# The ground ytterbium doublet in h-YbMnO$_3$ and the related low-temperature peculiarities of the compound

S.A. Klimin[1,*], N.D. Molchanova[1], N.N. Kuzmin[1], E.S. Sektarov[1], Lihua Yin[2], and M.N. Popova[1]

[1] *Institute of Spectroscopy, Russian Academy of Sciences, Troitsk, Moscow 108840, Russia*
[2] *Key Laboratory of Materials Physics, Institute of Solid State Physics, HFIPS, Chinese Academy of Sciences, Hefei 230031, People's Republic of China*

**Abstract**

We have performed detailed temperature-dependent study of optical *f-f* transitions of the Yb$^{3+}$ ions in h-YbMnO$_3$ by means of Fourier-transform spectroscopy. The splitting of the ground Kramers doublet as a function of temperature, $\Delta_0(T)$, for the Yb$^{3+}$ ion at 4b site was determined. The $\Delta_0(T)$ function follows the dynamics of the manganese magnetic moment below $T_N$ = 87 K, indicating, that the ytterbium subsystem is magnetized by the magnetic field generated by an ordered manganese subsystem, which is consistent with the results of previous studies. Excitation of the upper component of the split ground doublet plays a significant role in low-temperature dynamics of the h-YbMnO$_3$ crystal. Using the $\Delta_0(T)$ function we calculated the temperature behavior of the of the Yb(4b) magnetic moment: it is in clear agreement with the neutron data [*Phys. Rev. B* **98**, 134413, 2018]. The calculated contribution of Yb(4b) to heat capacity definitely explains the origin of the Schottky anomaly in the $C_P(T)$ dependence. A scenario for phase transitions in h-YbMnO$_3$ is proposed in which the energy gain in the ytterbium system plays a key role.

**Keywords:** h-YbMnO$_3$, hexagonal manganites, optical spectroscopy, *f-f* transitions, Kramers doublets, ground doublet model

* Corresponding author klimin@isan.troitsk.ru

# I. INTRODUCTION

Rare earth (RE, $R$) manganites $R$MnO$_3$ and their derivatives possess prominent physical properties and are among the most extensively studied functional oxides. The remarkable phenomena observed in these compounds, such as colossal magnetoresistance [1–3], giant magneto-elastic [4] and magneto-electric [5] couplings, significant magnetocaloric effect, ferroelectricity [6,7], multiferroicity [8,9], electromagnons [10], intricate magnetic phase transitions [11–14] make $R$MnO$_3$ manganites potentially attractive for the use as magnetoelectric devices, multiferroic memory elements, magnetic field and THz sensors, for antiferromagnetic (AFM) tomography [15], refrigeration [16], tunable microwave [17] and THz devices, UV photodetectors [18], ultrafast spintronic switching via electromagnons, and fundamental research platform.

$R$MnO$_3$ compounds can crystallize in two different crystal structures: orthorhombic (o-$R$MnO$_3$) and hexagonal (h-$R$MnO$_3$), both of which are of interest for both fundamental physics and practical applications. Hexagonal $P6_3cm$ structure of h-$R$MnO$_3$ is layered. Manganese is arranged in layers, forming a nearly ideal triangular 2D sublattice [19]. This leads to the magnetic frustration. In YMnO$_3$ with single magnetic subsystem, the 120° magnetic structure is realized [20] at temperatures $T<T_N$ [21,22]. A similar 120° magnetic structure of the manganese subsystem is realized in h-$R$MnO$_3$ with magnetic RE element $R$. The presence of a RE magnetic subsystem leads to complicated inter-layer interactions. $R^{3+}$ ions occupy two different crystallographic Wyckoff sites, $4b$ and $2a$, further we use Yb(4b) and Yb(2a) notations. Thus, the magnetic Yb$^{3+}$ ions form two magnetic sublattices.

In h-YbMnO$_3$, manganese subsystem orders in AFM phase at $T_N$ in the range of 75 - 90 K depending on the sample. The transition from the paramagnetic (PM) state to the AFM state has been confirmed by many different experiments, among them: magnetic susceptibility [23–27], magnetization [28,29], heat capacity [9,30], dielectric permittivity [9,25,28], second-harmonic generation [11], neutron scattering [26,31], Raman [32] and infrared (IR) [33,34] spectroscopies. Ordering at close temperatures occurs also in some other members of the h-$R$MnO$_3$ family, namely in compounds with R=Ho, Er, Dy, Tm [21,23]. At very low temperature in the range of 3-5 K, another feature arises, associated, according to the literature data, with the ordering of the Yb(2a) subsystem [26,28,30,35,36].

Previous spectroscopic study of $f$-$f$ transitions in YbMnO$_3$ revealed a splitting of the Yb(4b) Kramers doublets at $T_N$ [33]. This confirms the magnetic ordering, since the Kramers degeneracy can only be lifted by a magnetic field. Such a field appears in a crystal at $T<T_N$ in the form of an effective field $B_{eff}$ of the manganese magnetic subsystem acting on ytterbium. Transition of the $Yb^{3+}$ ion to the upper state of the split ground Kramers doublet can be considered as a low-energy CF excitation which influences the low temperature dynamics of a crystal. Energy of such excitation can be found by optical spectroscopy and can be used both to model low-temperature physical properties of the substance [37–44], and to clarify the nature of magnetic interactions. To perform such a calculation in the case of the title crystal one need to know the $\Delta_o(T)$ function, i.e., the temperature dependence of the splitting of the ground $Yb^{3+}$ Kramers doublet. Ref. [33] reports only splitting of the spectral lines, but gives no data on $\Delta_o$ and its temperature dependence. Two studies report on the value of $\Delta_o$ at low temperatures: (1) the heat capacity experiment [30] indicate the existence of a low-temperature excitation near 1.6 meV ($\approx 13$ cm$^{-1}$); (2) in the far IR spectra, a low-energy crystal-field (CF) excitation in YbMnO3 was detected 11.5 cm$^{-1}$ (4.5 K) [36].

Temperature behavior of the Yb(4b) magnetic moments was interpreted using phenomenological equations [26]. Good approximation of experimental data on Yb(4b) magnetic moments was achieved. Instead of phenomenological approach, the spectroscopic data can be directly used in the mean-field theory, describing magnetic interactions in mixed *d-f* magnetic systems [37,38,40,44].

In this paper we have performed a detailed study of temperature-dependent of transmission spectra in the region of f-f transitions of the $Yb^{3+}$ ion in h-YbMnO$_3$, using the method of high-resolution Fourier spectroscopy. We derived $\Delta_o(T)$ function and calculated (i) the effective magnetic field acting on Yb(4b) site and mean-field constant, (ii) temperature dependence of the magnetic moment $m_{Yb(4b)}$ and (iii) the contribution of ytterbium to heat capacity. We emphasize that the magnetic anisotropy of RE ions plays a significant role in the specificity of magnetic phase transitions in the family of hexagonal RE manganites.

## II. EXPERIMENT

The hexagonal YbMnO$_3$ single crystals were grown by the flux method as reported previously [45]. The stoichiometric initial materials of $Y_2O_3$ and $MnO_2$ were weighed and mixed with the flux $Bi_2O_3$ in a suitable ratio, placed into a platinum crucible and covered with a lid, and heated at 1200 °C for 20 h. Then the mixture was cooled slowly to 1000 °C at a rate of 3 °C/h, and finally the furnace was turned off. The as-grown crystals are of thin plate-like shape with an in-plane size of several millimeters and a thickness of about ~80 μm, as shown in Fig. 1a. The chemical composition of the as-grown crystals was analyzed by the energy dispersive spectroscopy (EDS) technique, which showed a ratio of Yb:Mn~1:1 in the crystals.

Transmission spectra were recorded with a resolution up to 0.2 cm$^{-1}$ using a Bruker IFS 125 HR Fourier transform spectrometer. A Sumitomo RP-082E2S closed-cycle helium cryostat was used for cooling the samples. To reduce thermal loads, a double-walled polished screen with openings for input and transmitted light was connected to the first cooling stage. To ensure good heat transfer to the cryostat, the sample was glued with silver paste. Temperature was measured with a Lakeshore DT-670 silicon diode, mounted close to the sample, using a Lakeshore 335 temperature controller with an accuracy of ±0.05 K.

## III. RESULTS AND DISCUSSION

### A. Crystal structure of h-YbMnO$_3$

The crystal structure of h-YbMnO$_3$ belongs to the hexagonal *P6$_3$cm* group [19,46–50]. Fragments of the structure, generated with the help of the *Balls&Sticks* program [51], are shown in Figs. 1b and 1c. Manganese *ab* layers are shown both as perspective view (Fig. 1b) and as a 2D triangle lattice (Fig. 1c). The minimal distance between Mn$^{3+}$ ions is 3.52 Å in the *ab* plane and 6.18 Å in closest planes. Such a geometry predetermines the properties of the magnetic interactions, namely, strong frustration and weak interplane connection. Ytterbium ions form another *ab* layers alternating with manganese ones. In Fig. 1c, they are shown as a net of YbO$_7$ polyhedra sharing common edges and filling the entire space of the plane. Two types of YbO$_7$ polyhedra formed by ytterbium ions in 2a and 4b positions, are painted blue and light green, respectively. They are shown in more detail in Fig. 1b. Both polyhedra types have similar view, they can be described, more or less, as capped triangular antiprisms. 2a (4b) polyhedron has C$_{3v}$ (C$_3$) symmetry. The dashed lines connect six manganese sites, closest to Yb(4b), with oxygen atoms of the YbO$_7$ polyhedron, i.e., they show segments of the Mn-O-Yb superexchange paths.

### B. Transmission spectra of h-YbMnO$_3$

The transmission spectra measured at several temperatures are shown in Fig. 2. Given spectral region includes the $^2F_{5/2}$ multiplet of the Yb$^{3+}$ ion and broad bands of the Mn$^{3+}$ spectrum. All the spectral lines narrow at cooling. As a result, Yb multiplet hardly seen at room temperature becomes nicely registerable at low temperatures. This is due to the opening of transparency window when Mn bands narrow.

The ytterbium ion has very simple energy-level scheme of 4*f* electron states. Yb$^{3+}$ has the $^4f^{13}$ electron configuration with one hole in the 4*f* shell. This gives just two multiplets (the ground $^2F_{7/2}$ and the excited $^2F_{5/2}$) resulting from the spin–orbit coupling. In a CF with lower than cubic symmetry, these multiplets split into four and three Kramers doublets, respectively, as is shown in Fig. 2b. No one can find a simpler scheme among the trivalent RE ions.

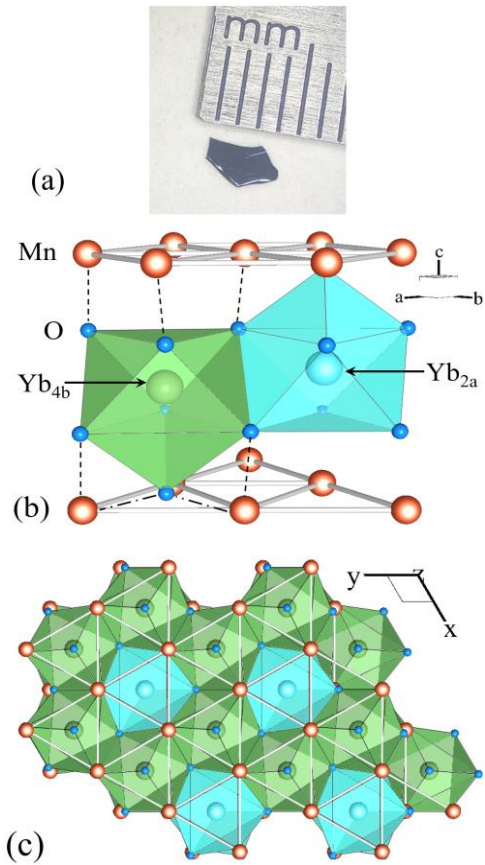

Fig. 1. (a) Photo of the sample used in the experiment and (b,c) two fragments of the crystal structure of h-YbMnO$_3$, namely, (b) a perspective view perpendicular to the $c$ axis: succeeding manganese 2D triangular layers with YbO$_7$ polyhedra of two types between them; dotted lines show superexchange paths from manganese atoms to ytterbium via oxygens, and (c) a view along the $c$ axis.

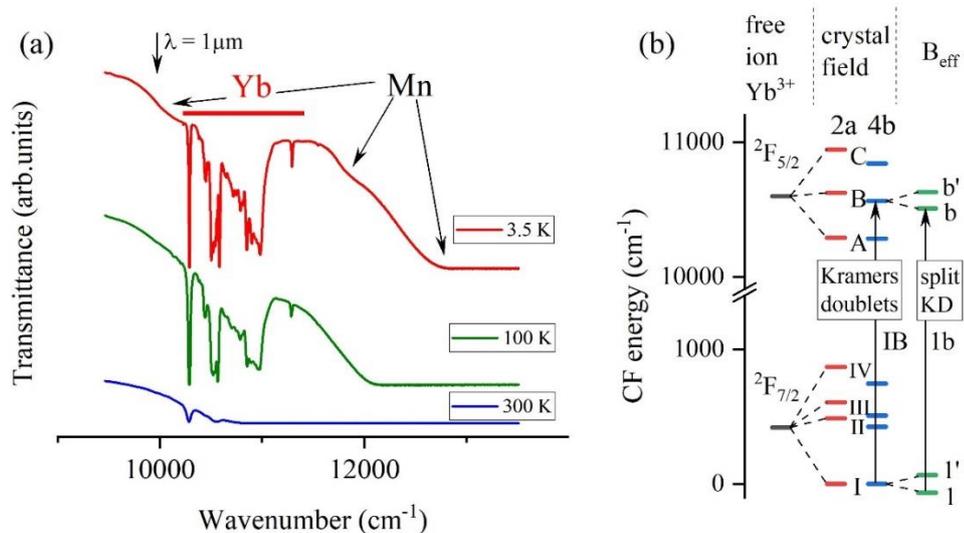

Fig. 2. (full width) (a) Transmission spectra of YbMnO$_3$ at several temperatures over a wide spectral range, including the $f$-$f$ transition in the Yb$^{3+}$ ion. (b) Yb$^{3+}$ CF level diagram. CF energies are taken from Ref. [33]

At low temperatures, when only the ground level "1" is populated, each ytterbium site gives three spectral lines. For two crystallographic ytterbium sites in YbMnO$_3$, one can observe maximum six spectral transitions. However, the spectrum at 3 K, shown in Fig. 2a, is much more complex. Such a picture is typical for Yb spectrum in a crystal [52–56]. The reason lies in the specificity of Yb intermultiplet transitions. One can note that the so called "0-0" transition (denoted as 1A in the scheme of Fig. 2b and corresponding to the line 10 290 cm$^{-1}$ in the spectrum) is much more intensive than the rest of the spectrum, the line is strongly saturated. The large cross-section of 0-0 transitions in Yb$^{3+}$ ion is a reason of using ytterbium as a sensitizer [57,58]. Excluding 1A line, the rest of the spectrum at low temperatures consists of comparable in intensities CF transitions (1B and 1C) and reach spectra of vibronics.

A detailed study of the dependence of the intermultiplet transition on temperature reveals that the most noticeable changes occur at temperatures close to $T_N$, see Fig. 3. The most impressive is a clearly visible splitting of the spectral line at 10566 cm$^{-1}$ exactly at $T_N$. At 87 K, a magnetic phase transition occurs in the crystal. The manganese magnetic subsystem becomes ordered. A magnetic field $B_{eff}$ arises in the crystal, lifting the Kramers degeneracy.

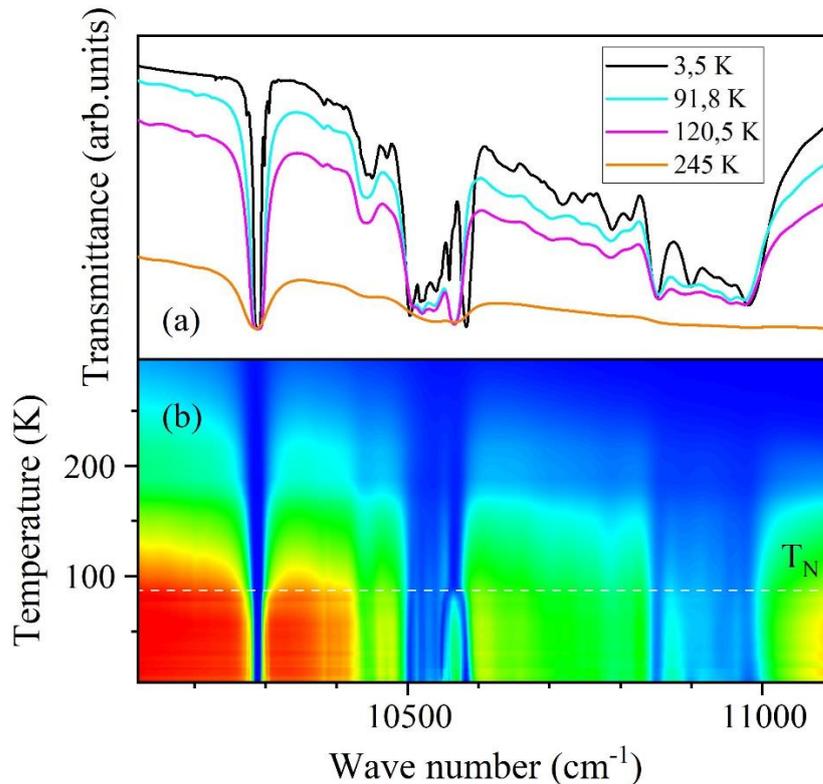

Fig. 3. (a) Intermultiplet $^2F_{7/2} \rightarrow {}^2F_{5/2}$ transition in the Yb$^{3+}$ ion in YbMnO$_3$ at different temperatures. (b) Transmission intensity map in frequency-temperature coordinates.

## C. $\Delta_0(T)$ function for Yb(4b) site in YbMnO$_3$

The splitting of the spectral line at 10566 cm$^{-1}$ is shown in more details in Fig. 4a. According to the scheme of Fig. 4b, in the general case, one can observe four spectral lines in the AFM state instead of one in the PM phase. This is just the case of the line 10566 cm$^{-1}$. Curved dashed arrows in Fig. 4a depict four spectral lines arising due to the splitting of Kramers doublets of both initial and final levels of the transition. Two of the lines, namely 1′a and 1′a′, both corresponding to the transitions from the upper component of the split ground Kramers doublet, "freeze" at low temperatures. This is due to the depopulation of the 1′ state. Instead, the intensity of 1a and 1a′ lines grows with cooling. The reason is the same – the change of the population of the initial state of transition, namely, the population of the 1 state grows with cooling. The distances between four components immediately allows one to find the splittings of both ground and excited Kramers doublets. At the lowest (in this work) temperature the values of this splittings are $\Delta_0(3.5\ K) = 11$ cm$^{-1}$ and $\Delta_A(3.5\ K) = 22$ cm$^{-1}$. The obtained value of $\Delta_0$ is in agreement with assumption of [36], where the 11.5 cm$^{-1}$ line was registered in THz experiment.

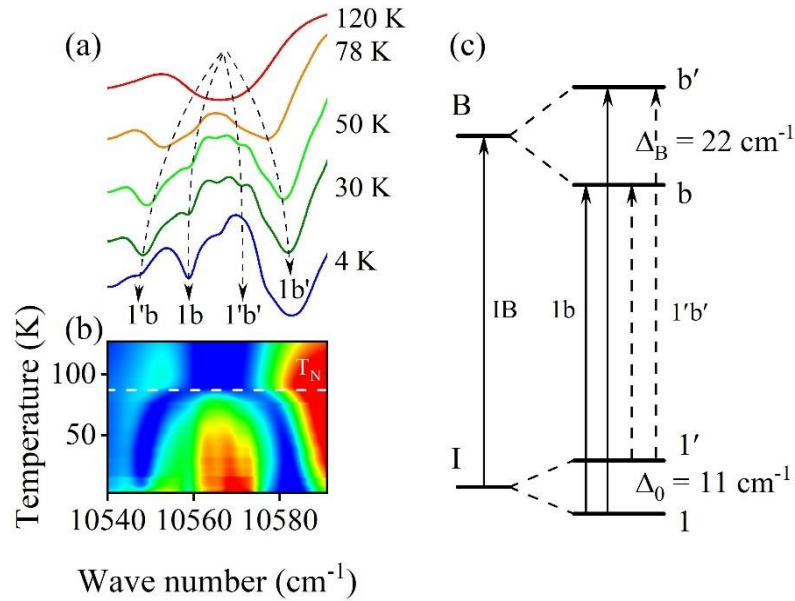

Fig. 4. Splitting of the 10566 cm$^{-1}$ spectral line shown as (a) spectra at different temperature and (b) color intensity map. (c) Energy level diagram. Splitting of the ground and excited Kramers doublet participating in optical transition.

Further confirmation of the value 11 cm$^{-1}$ for $\Delta_0(3.5\ K)$ follows from the temperature dependence of the integral intensity of the line 1b. Fig. 5a shows how the intensity of the line 1b depends on temperature in comparison with the population of the level 1′. The experimental points follow the Boltzmann population $n=\exp(-\Delta_0(T)/kT)/(1+\exp(-\Delta_0(T)/kT))$ depicted in Fig. 5a for $\Delta_0(3.5\ K) = 11$ cm$^{-1}$.

From the temperature dependences of the positions of the spectral components we have find a $\Delta_0(T)$ function which is shown in Fig. 5b by orange circles. The high-temperature part of the $\Delta_0(T)$ function was obtained by processing the line's half-width, which begins to increase at temperatures well above $T_N$. The increase in the half-width is due to the onset of splitting of the Kramers doublets as a result of the short-range correlations in the manganese subsystem. Thus, $\Delta_0(T)$ function has a "tail" which extends into the region $T<T_N$. Such a tail, as a result of magnetic correlations, is characteristic of the low-dimensional systems [59]. In our case, this is the 2D subsystem of manganese in h-YbMnO$_3$.

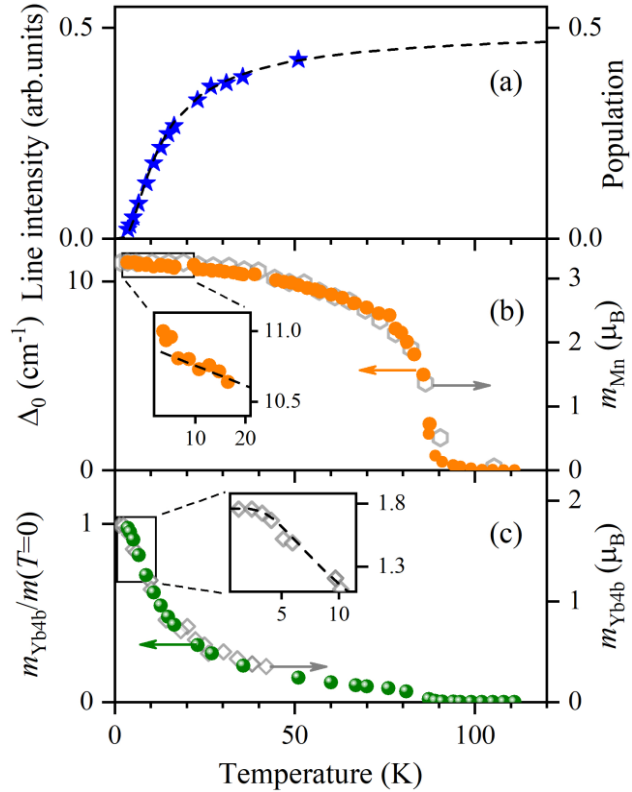

Fig. 5. Temperature dependences of several physical parameters of YbMnO$_3$. (a) Integral intensity of the 1′b line at 10547 cm$^{-1}$ (blue stars) and the population of the 1′ level calculated using Eq. 4 with $\Delta_0(0)=11$ см$^{-1}$ (dotted line). (b) The splitting $\Delta_0$ of the Yb(4b) ground doublet (orange circles) and the manganese magnetic moment $m_{Mn}$ [31] (grey circles). (c) Magnetic moment of ytterbium in 4b site $m_{Yb(4b)}$ from neutron experiment [26] (grey rhombs) and calculated using our spectroscopic data (green circles). Inserts in (b) and (c) show selected data on a larger scale.

### D. Calculations with $\Delta_0(T)$ function

The $\Delta_0(T)$ function is an important physical quantity. In many cases, it allows to explain low-temperature dynamics of the crystal. First, we note that the $\Delta_0(T)$ function is almost proportional to the temperature-dependent value of the magnetic moment $m_{Mn}$ found in neutron experiment [31] and shown in Fig 2b by grey circles. This observation clearly leads to the conclusion that ytterbium in the 4b site is magnetized by the effective field acting on Yb and created by the Mn subsystem. Pay attention, that the value of $T_N \approx 80$ K was reported in [31], in contrast to the value $T_N \approx 87$ K which we report here. As it was mentioned above, the temperature of magnetic ordering of h-YbMnO$_3$ depends noticeably on the sample. The reason for that can lie in the quality of the samples, which is also typical for other types of compounds, see, e.g., [60]. To compare the data obtained in our work and in Ref. [31], a *T*-scale adjustment with a coefficient of 87/80 was introduced.

The magnitude of the effective magnetic field acting on the Yb(4b) site can be evaluated form the simple relation

$$\Delta_0 = g_{Yb}\mu_B B_{eff}, \qquad (1)$$

where $g_{Yb}$ and $\mu_B$ are g-factor of the Yb$^{3+}$ ion and the Bohr magneton, respectively.

The g-factor of Yb$^{3+}$ ion in position 4b is known from neutron experiments on h-YbMnO$_3$. The values of $m_{Yb(4b)} = 1.77$ µ$_B$ at 1.5 K [26] and 1.76 µ$_B$ at 2 K [31] coincide within the experimental error. Note that, at $T < 3$ K temperature dependence of the magnetic moment of Yb(4b) shows a plato, see insert to Fig. 5c. Using the relation $g_{Yb} = 2m_{Yb}(T=0)/\mu_B$ we were able to evaluate the B$_{eff}$, which turned out to be equal to 7 T. The Eq. 1 is a simplified version and is used only to estimate the value of $B_{eff}$. The exact relation takes into account the anisotropy of the RE ion, which will be discussed below.

From the other side, the value of the mean-field constant also can be calculated according to the relation:

$$B_{eff} = \lambda_1 m_{Mn}. \qquad (2)$$

Using the values of $m_{Mn} = 3.23$ µ$_B$ at 1.5 K [26] and $m_{Mn} = 3.41$ µ$_B$ at 2 K [31], which are relatively close to each other, we obtain two values for $\lambda_1$, namely, 2.17 and 2.05 T/µ$_B$, respectively. Average of two values is almost equal to the estimation $\lambda_1 = 2.1$ T/µ$_B$ from Ref. [26].

The temperature dependence of the Yb$^{3+}$ magnetic moment can be described in the frame of the mean-field theory with phenomenological parameter $\lambda_1$:

$$m_{Yb}(T) = \frac{1}{2}g_{Yb}\mu_B tanh\frac{g_{Yb}\mu_B \lambda_1 m_{Mn}(T)}{2kT}. \qquad (3)$$

Such analysis was made for h-YbMnO$_3$ [26]. The experimental neutron data on $m_{Yb(4b)}(T)$ dependence were nicely described by Eq. 3. The authors argue, that "The unusual upward curvature comes from the *tanh* function", with providing no justification for the application of this function. In a study of magnetic behavior of another *d-f* compound, Nd$_2$BaNiO$_5$, the authors apply successfully *tanh* function as "purely empirical" [61] for description of the temperature behavior of the neodymium and the nickel magnetic moments. The same *tanh* function was used in Ref. [62] to describe similar relationships in the mixed magnetic *d-f* system Er$_2$BaNiO$_5$. In [62], the

*tanh* relation was derived as a consequence of the statement that "only the ground doublet can be considered to contribute to the magnetic moment of erbium subsystem µEr(*T*)". Here we show once again, but in more detail, how the *tanh* function arises when considering the "ground-doublet model". One just has to accept the validity of the statement that the magnetic moment of the Kramers $Yb^{3+}$ ion is proportional to the difference in populations of split sublevels of the ground Kramers doublet. This statement is true as the wavefunctions of sublevels correspond to the states with oppositely directed magnetic moments. In this case,

$$m_{Yb} = m(0)_{Yb}(n_1 - n_{1'}), \qquad (4)$$

$$n_{1'} = n_1 e^{(-\Delta_0/kT)}, \qquad (5)$$

$$n_{1'} + n_1 = 1, \qquad (6)$$

where $n_{1'}$ and $n_1$ are the populations of the sublevels of the split ground Kramers doublet, and $m_{Yb(4b)}(0)$ is magnetic moment of the Yb(4b) ion at the temperature close to zero. The set of Eqs. 4-6 we call "the ground-doublet model". Simple calculations using Eqs. 4-6 explicitly show the appearance of the hyperbolic tangent function and give the following result:

$$m_{Yb}(T) = \frac{1}{2} g_{Yb} \mu_B \tanh \frac{\Delta_0(T)}{2kT}. \qquad (7)$$

This equation is completely equivalent to Eq. 3: using Eqs. 1 and 2, one can easily derive the relation $\Delta_0 = g_{Yb}\mu_B\lambda_1 m_{Mn}$. The difference in using Eqs. 3 and 7 is that Eq. 7 has no phenomenological parameters and someone can take directly experimental data on $\Delta_0(T)$ to calculate the $m_{Yb(4b)}(T)$ dependence. Ratio $m_{Yb(4b)}(T)/ m_{Yb(4b)}(0)$ is shown in Fig. 5c together with the experimental data on $m_{Yb(4b)}(T)$ from neutron experiments [31]. The good agreement between the two data sets confirms the correctness of the ground doublet model and highlights the importance of optical methods in studying magnetic crystals.

The temperature change of the RE magnetic moment in the field of an ordered *d* subsystem, characterized in the case of h-YbMnO$_3$ by a sharp increase at temperatures below 15 K, is sometimes called intriguing or a consequence of the *tanh* function in the phenomenological equations of the mean-field model. Looking more deeply, we can say that this type of dependence is a consequence of the ground-doublet model (see Equations 4-6). The magnetic moment of a RE ion is proportional to the population difference between the sublevels of the split ground doublet.

### E. Schottky anomaly in the heat capacity of h-YbMnO$_3$

Any energy state in a crystal becomes populated upon heating. An additional portion of energy d*E* is required and it contributes to heat capacity $C_P=dE/dT$. The most demonstrative is the so-called Schottky anomaly. It arises at low temperature when the lattice contribution almost vanishes, but low-energy excitations exist in a crystal. Such a situation frequently happens in crystals containing rare earths, see, e.g., Refs. [14,37,63,64]. The heat capacity of h-YbMnO$_3$ has several singularities [9,30], namely, two λ-type anomalies, one at $T_N$ and another at $T_R$, and the Schottky anomaly with a maximum at ~7.5 K (see insert to Fig. 6). Knowing the $\Delta_0(T)$ function it is easy to calculate a contribution of ytterbium into $C_P$ of h-YbMnO$_3$ using the following relation for the two-level system (see, e.g., Ref. [37,65]):

$$C_{P,Yb(4b)} = n_{Sch} x R \left(\frac{\Delta_0(T)}{kT}\right)^2 \frac{e^{\Delta_0(T)/kT}}{(1+e^{\Delta_0(T)/kT})^2}. \tag{8}$$

Here, $R$ is the gas constant, $x$ is the number of Yb atoms in a formula unit, $n_{Sch}$ is the correction factor. As we calculate the contribution of ytterbium 4b subsystem, we set $x$ equal to 2/3. If it were Yb(2a), $x$ would be one third. Factor $n_{Sch}$ accounts for the effective number of ytterbium ions contributing to heat capacity. For an ideal crystal, $n_{Sch}$=1. In a real crystal it can differ from unity.

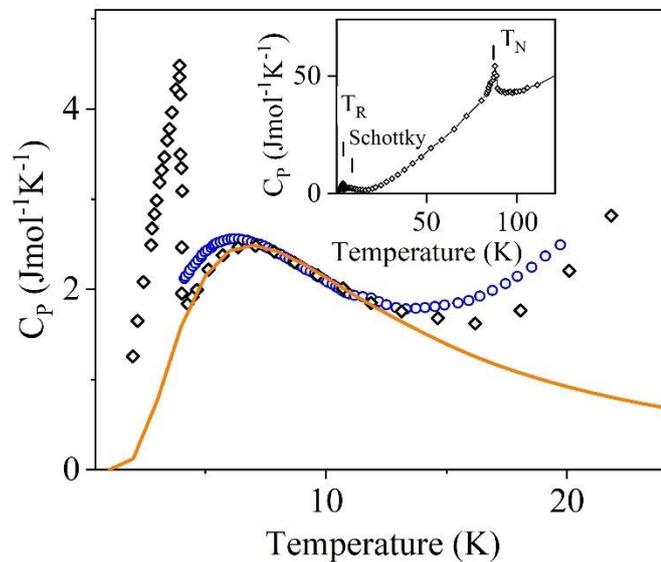

Fig 6. Heat capacity of h-YbMnO$_3$. The calculated ytterbium contribution (orange solid line) compared to experimental data from Ref. [9] (black open rhombs) and from Ref. [30] (blue open circles). Insert: experimental data [9] in a wide temperature range.

Equation 8 implies that the other CF levels of the ytterbium ground multiplet are unoccupied. The absence of "hot lines" in the low-frequency region from 0-0 line even at room temperature enable us to conclude that the energy of the first excited level (II) of the ground $^2F_{7/2}$ multiplet is rather high. This conclusion is in agreement with Ref. [33]. The CF calculations [33] give the energies of 429 cm$^{-1}$ and 487 cm$^{-1}$ for the level "II" for Yb(4b) and Yb(2a), respectively. So, there is a large gap between the ground and the first excited Kramers doublet, which clearly implies that at low temperatures only the ground Kramers doublet is populated and Eq. 8 applies.

The result of calculation using Eq. 8 is shown in Fig. 6 by orange solid line. Here it is compared with the experimental data taken from two datasets [9,30]. Data from two experiments are close to each other. Visible difference can lie in both temperature measurement error or specificity of a samples, which are not ideal. The calculated curve fits well the data represented by black diamonds [9]. An interesting conclusion arising from the good fit is the confirmation that the line 10566 cm$^{-1}$ belongs to the Yb(4b) site, for which $x=2/3$. Otherwise, for the Yb(2a) position, $x=1/3$ would have to be used, and the height of the calculated anomaly would be insufficient to describe the experimental data. Note that for a good match we had to use a coefficient $n_{Sch}$ of 0.93. As was mentioned, this means that h-YbMnO$_3$ crystals used in heat capacity experiments are not ideal. Nonideality may be connected with the presence of other chemical phases (e.g., o-YbMnO$_3$), impurities, non-stoichiometry, partial changing in manganese valency, etc. The wide range of the Neel temperature observed in different experiments can also be attributed to the specificity of the samples and their different quality.

Note that the maximum value of $C_P$ at Schottky anomaly in experimental set shown by blue circles is slightly higher. To obtain the same value in fitted maximum, we need to set the value of $n_{Sch}$ to 0.96. This may mean, if no experimental error, that the crystal used in experiment [30] has fewer defects than that in Ref [9]. In any case, the opportunity is opened to use the maximum value in Schottky anomaly as a criterion of crystal quality. This statement is supported by the study of Ca-doped YbMnO$_3$, which clearly demonstrates a decrease in the amplitude of the Schottky anomaly with increasing $x$ in the Yb$_{1-x}$Ca$_x$MnO$_3$ composition [30].

### F. The scenario of magnetic phase transitions in h-YbMnO$_3$

Magnetic ordering and intriguing phase transitions in RE h-manganates were studied for decades since their discovery in 1963 [20]. Although certain facts have been established, there are also many contradictions. Moreover, the discussion of the issue is complicated by details related to the presence of domains, magnetoelectric effects, magnetostriction and symmetry-lowering displacement of atoms. Our study does not claim to be definitive, but it does draw attention to a specific process, involving ytterbium ground doublet that has not been taken into account in previous studies.

According to the adopted hierarchy of interactions in mixed *d-f* magnetic systems, Mn-Mn one is the strongest and it is self-sufficient for ordering of the *d* magnetic subsystem. However, the magnetic structure of ordered manganese, which occurs at temperature $T_N$, differs for different *R* in h-*R*MnO$_3$ manganites. Always 120° frustrated, it is distinguished by the spatial orientation of the magnetic moments relative to the crystal axes. Four acceptable symmetry types are allowed. Moreover, depending on the sample, the manganese magnetic moments can undergo a spin-orientation phase transition with temperature. Can RE ions play a significant role in the realization of a particular type of magnetic structure in the *d* subsystem? Our answer is yes. The accepted explanation in the literature suggests that the next strongest type of magnetic interaction (after Mn-Mn), namely, R-Mn, leads to the polarization of the RE magnetic moments in the magnetic field generated by the manganese subsystem, while the weakest *R-R* interactions can only be realized at low temperatures, leading to the second Néel temperature. While this logic is undeniable, it is highly mechanistic and does not reveal the nature of phase transitions. Moreover, due to its limitations, it could lead to incorrect conclusions in some cases.

The most important point is that this logic does not take into account the single-ion magnetic anisotropy of the RE ion. Our previous studies show that in mixed *d-f* magnetic compounds RE can dictate the type of magnetic structure [55,66–68]. The reason is somewhat trivial. RE ground-doublet splitting leads to a decrease in the system's energy, but such splitting is only possible with a certain type of magnetic structure of the *d* subsystem. This type may be energetically unfavorable. Within the pure manganese sublattice there may be a magnetic structure with lower energy, but the energy gain due to the RE ground-doublet splitting can be larger. Thus, it is the single-ion anisotropy of the *f* ion that can cause a rearrangement of the magnetic moments of manganese.

Let us return to promised relation for $\Delta_0$ taking into account the RE anisotropy:

$$\Delta_0(T) = \mu_B \sqrt{\sum (g_i B_{\text{eff},i}(T))^2} \qquad (9)$$

where $g_i$ and $B_{\text{eff},i}$ are the *i*-th components ($i = x, y, z$) of the ytterbium *g*-factor and of the effective magnetic field acting on ytterbium. The *g*-factor of $RE^{3+}$ ion can be highly anisotropic as a consequence of CF effects. Thus, we say "easy-axis" anisotropy if $g_z \neq 0$, $g_x = g_y = 0$, and "easy-plane" if $g_z = 0$, $g_x, g_y \neq 0$. Let us highlight three important conclusions that follow from Eq. 9.

(1) Not every exchange field splits the ground doublet. For example, in the case of easy-axis anisotropy, a field directed perpendicular to the *c* axis will not cause splitting and, as a consequence, appearance of the RE magnetic moment.

(2) Due to splitting, the ground state of the RE ion is lowered, giving the system an energy gain. If this gain is large enough, it can promote energy-consuming spin flips in the *d* subsystem.

(3) The value of $\Delta_0$ depends on temperature. Energy characteristics associated with splitting can cause certain sharp processes at critical temperatures.

The energy gain $\Delta\mathcal{E}$ due to the lowering of the RE ground state can be evaluated as $1/2\Delta_0$ per one atom, or per mol it reads:

$$\Delta\mathcal{E}(T) = \frac{1}{2} x N_A h c \Delta_0(T)(n_1 - n_{1'}), \qquad (10)$$

where *x* is the same as in Eq. 8. For Yb(4b) with $\Delta_0 = 11$ cm$^{-1}$ at low temperatures when level 1′ is empty, $\Delta\mathcal{E} = 437$ Jmol$^{-1}$ which is rather large value and have to be taken into account considering the type of magnetic structure of the ordered manganese subsystem. All present theories considering the magnetic structure of manganese subsystem neglect Mn-Yb interaction referring to the hierarchy of interactions. However, based on Eq. 9, it can be concluded that it is the Mn-Yb interaction that can lead to a large energy gain.

According to neutron data, the magnetic moments in both ytterbium subsystems are directed along the *c* axis [31]. In this case, we can speak of easy-axis anisotropy for both Yb(4b) and Yb(2a) centers. It is known that the anisotropy of the rare-earth ion is determined by the crystal field. Qualitatively, although the symmetry of the two ytterbium centers differs, the difference is small (see Fig. 1), which can lead to similar effects determined by the crystal field. Surprisingly, despite this geometry "closeness", the ground doublet of the Yb(4b) center splits at $T_N$, but there is no spitting for the Yb(2a) center. The exchange field on both ytterbium sites can only arise due to the antisymmetric Dzyaloshinsky-Moriya (DM) exchange [26]. The symmetric exchange for both centers is canceled out due to the geometry. The critical difference in the exchange field is surprising: for the center 2a the field is zero. It is known that the DM exchange is very sensitive to the Mn-O-Yb angles. This may be what determines the difference in the exchange fields for the two centers. However, at zero DM exchange, at temperatures below 3.5 K, the magnetic moments of the Yb(2a) center are ordered, organically implying a splitting of the ground state (Eq. 4).

Where does the exchange field at the Yb(2a) center come from? According to neutron studies, Yb(4b) ions orders ferromagnetically at this low-temperature transition. The exchange field at the Yb(2a) arises from neighboring Yb(4b) ions. At temperatures above 3.5 K, Yb(4b) ions order antiferromagnetically, and the average field at the Yb(2a) center is zero. Thus, the low-temperature phase transition is a spin-reorientation transition. According to neutron data, the magnetic structure of manganese differs for temperatures above and below 3.5 K [26,31].

To summarize the above, we propose the following interpretation of the magnetic phase transitions in h-YbMnO$_3$. The diagram in Fig. 6a shows the relative changes in the energy of three magnetic subsystems (Mn, Yb(4b), and Yb(2a)) of h-YbMnO$_3$ manganite. The energy curves for both Yb sites are drown on the basis of spectroscopic data obtained in this study. We also plotted three of the four possible [11] levels for the energy of manganese magnetic subsystem. In the temperature range slightly above $T_N$, a short-range magnetic correlations arise. Short-range magnetic correlations of the manganese sublattice create a dynamic splitting of the ground state of ytterbium at the 4b position. As discussed above, the field at the Yb(2a) center is zero, and there is no splitting for this center. As a result, the energy of the ground state of ytterbium at the 4b position decreases. In this case, manganese may not be at the energy minimum; its energy may increase, as shown in the diagram. However, manganese creates the field necessary to split the ground state of ytterbium, so the total energy of the system does not grow ($\Delta_1^{Mn}+\Delta_1^{Yb(4b)} \leq 0$, see Fig. 7a). At temperature $T_N$, a transition to the antiferromagnetic state AFM1 occurs. As the temperature decreases further, the splitting of the ytterbium 4b ground state doublet increases, and the energy of the system decreases due to the Yb(4b) center.

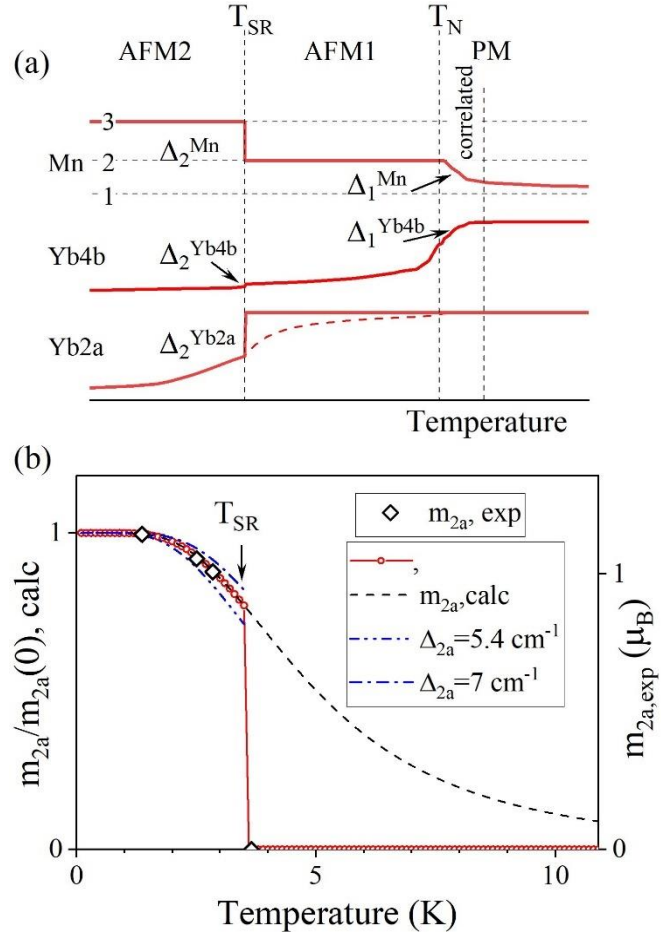

Fig. 7. (a) Schematic temperature dependences of relative energies of the three magnetic subsystems (Mn, Yb(4b), Yb(2a)) of h-YbMnO$_3$. (b) Experimental Yb(2a) magnetic moments (rhombs) [26] compared to modeling (lines), see text.

At 3.5 K, a spin-reorientation phase transition occurs. An effective field appears at the Yb(2a) center, splitting its ground-state Kramers doublet. As noted above, the field from manganese is always zero at the Yb(2a) center, due to geometric properties. However, the moments of the Yb(4b) center become ferromagnetically ordered. They create a field that splits the ground-state doublet of the Yb(2a) center. As soon as the energy gain due to lowering the Yb(2a) ground state becomes sufficiently large, an energy-consuming spin flip of the other two subsystems occurs. During this transition, the splitting of the Yb(4b) doublet can also change: in the insert of Fig. 5b, small changes in $\Delta_0^{Yb(4b)}$ are visible when approaching $T_{SR}$. Above the $T_{SR}$ temperature, the transition is impossible, since the energy gain is insufficient for the spin flip ($\Delta_2^{Mn}+\Delta_2^{Yb(4b)}+\Delta_2^{Yb(2a)} > 0$). Below the $T_{SR}$ temperature, the energy of the system decreases due to the Yb(2a) center ($\Delta_2^{Mn}+\Delta_2^{Yb(4b)}+\Delta_2^{Yb(2a)} < 0$).

Using Eq. 7, we were able to calculate how the magnetic moment of ytterbium at position 2a changes with temperature. Similar to equations (1) and (2), we write the following equations:

$$\Delta_0^{Yb2a} = g_{Yb2a}\mu_B B_{eff}^{2a}, \tag{11}$$

$$B_{eff}^{2a} = \lambda_2 m_{Yb4b}, \tag{12}$$

from which it follows:

$$\Delta_0^{Yb2a}(T) = \Delta_{2a} m_{Yb4b}(T)/m_{Yb4b}(0). \tag{13}$$

The ratio $m_{Yb(4b)}(T)/m_{Yb(4b)}(0)$ was found earlier, see Fig. 5c. The factor $\Delta_{2a}$, i.e., the splitting of the Yb(2a) ground doublet at zero temperature, will be used as an adjustable parameter. Further, Eq. 7 was used to calculate the Yb(2a) magnetic moment. Just as in the case of calculating the moment of ytterbium at position 4b, we present the ratio $m_{Yb(2a)}(T)/m_{Yb(2a)}(0)$ in Fig. 7b.

The dashed line shows the calculated magnetic moment of Yb(2a) over a wide temperature range. However, the splitting described by Eq. 13 exists only in the temperature range $T<T_{SR}$, i.e., only when an effective field appears as a result of reorientation. In the range $T>T_{SR}$, we set $\Delta_0^{Yb(2a)}=0$ and $m_{Yb(2a)}=0$. The curve consisting of connected red circles satisfies the latter condition. It is evident that the calculated curve describes well the data on the magnetic moment $m_{Yb(2a)}$ [26], shown in the Fig. 7b by symbols. In this case, we used $\Delta_0^{Yb(2a)} = 6.2$ cm$^{-1}$. Other values do not provide such a good agreement with the experimental data. As an example, the blue dotted line in Fig. 7b shows the results of calculation for $\Delta_0^{Yb(2a)} = 5.5$ and $6.9$ cm$^{-1}$. Thus, this calculation procedure can be considered as a fitting that reveals the value of $\Delta_0^{Yb(2a)}$, i.e., the splitting of the Yb(2a) ground doublet at zero temperature.

## IV. CONCLUSION

We have studied temperature-dependent transmission spectra of h-YbMnO$_3$ crystal, one of the representatives of the family of RMnO$_3$ compounds with hexagonal structure, well-known due to their unique physical properties including intriguing magnetic phase transitions. Interplay between three magnetic subsystems, namely, frustrated manganese and two ytterbium (Yb(4b) and Yb(2a)) ones, produces changes in the magnetic structure. In h-YbMnO$_3$, in addition to AFM ordering at $T_N$, another phase transition occurs at rather low temperature ~ 3.5 K, attributed in the literature to the ordering of Yb(2a) subsystem. Our study was encouraged by previous spectroscopic paper by Diviš et al. [33] who demonstrated that some spectral lines of *f-f* transitions are split in AFM state. We continued to study this effect. First, we have shown that the *ground* Kramers doublet of Yb in 4b position is split. The splitting is proportional to the changing-with-temperature magnetic moment of manganese, which evidences that Yb(4b) ions are magnetized in the effective magnetic field arising in the crystal at the magnetic ordering of the *d* subsystem. Second, we derived how the splitting $\Delta_o$ depends on temperature, which allowed us to apply a model of ground doublet. Within this model, we have calculated the temperature dependence of the Yb(4b) magnetic moment and the Schottky anomaly in heat capacity, which turned out to be in good agreement with existing experimental data.

Analyzing our data together with literature information, we propose the following scenario of magnetic phase transitions in h-YbMnO$_3$. The magnetic ordering of the manganese subsystem happens due to the interaction within the subsystem. The ordering of Yb magnetic moments in 4b site occurs at $T_N$ as magnetization in the effective magnetic field generated by ordered manganese. This field is due to the antisymmetric DM exchange. It effectively acts on the Yb(4b) site and is zero at the Yb(2a) one. Energy gain due to the lowering of the ytterbium ground state is a driving motive for realizing a particular type of manganese magnetic structure. At $T_N$, the Yb(4b) center contributes to the lowering of the crystal energy. At $T_{SR} = 3.5$ K such contribution from Yb(2a) center promotes spin-reorientation transition. Both Yb(4b) and Mn subsystems change the type of ordering, and the former flips from AFM arrangement to the FM one. Each Yb(2a) ion, being surrounded by six Yb(4b) neighbors, experiences the action of symmetric exchange field, which splits its ground doublet. As a result, the Yb2a subsystem gets magnetized.

## ACKNOWLEDGEMENTS

The spectroscopy study was carried out within the framework of state assignments of the Ministry of Science and Higher Education of the Russian Federation for the Institute of Spectroscopy Nos. FFUU-2025-0004 (S.A.K., M.N.P.) and FFUU-2024-0004 (A.D.M., N.N.K., E.S.S.). The cooperation between the researchers from the Institute of Spectroscopy RAS and the Institute of Solid State Physics, HFIPS CAS is in the frame of Joint Research Agreement.